\documentclass[twocolumn,showpacs,preprintnumbers,amsmath,amssymb]{revtex4}
\usepackage{graphicx}

\newcommand{\be}{\begin{equation}}
\newcommand{\ee}{\end{equation}}
\newcommand{\ba}{\begin{eqnarray}}
\newcommand{\ea}{\end{eqnarray}}
\newcommand{\ds}{\displaystyle}
\newcommand{\MPl}{M_{\mathrm{Pl}}}
\newcommand{\MEW}{M_{\mathrm{EW}}}
\newcommand{\hs}{\hspace{-1.12mm}}
\newcommand{\eq}{\hs & = & \hs}
\newcommand{\eqpluss}{\hs & + & \hs}
\newcommand{\eqequiv}{\hs & \equiv & \hs}
\newcommand{\eqnl}{\hs & & \hs}
\newcommand{\eqto}{\hs & \to & \hs}
\newcommand{\vs}{\vspace{1.5mm}}
\newcommand{\addspace}[2][3mm]{\raisebox{0mm}[#1][#1]{$\displaystyle #2$}}
\newcommand{\addsmsp}[1]{\addspace[2mm]{#1}}
\newcommand{\boldk}{{\boldsymbol{k}}}
\newcommand{\boldx}{{\boldsymbol{x}}}
\newcommand{\ned}[3]{{#1}^{#2}_{\phantom{#2} #3}}

\begin{document}

\title{Lagrangian formalism of gravity in the Randall-Sundrum model}

\author{Petter Callin}
  \email{n.p.callin@fys.uio.no}
\author{Finn Ravndal}
  \email{finn.ravndal@fys.uio.no}
\affiliation{Department of Physics, University of Oslo, N-0316 Oslo, Norway\\}

\date{October 3, 2005}

\begin{abstract}
We derive the effective Lagrangian of the physical four-dimensional fields in
the Randall-Sundrum (RS) model, and use this to calculate the Newtonian
gravitational potential between two point sources on the brane. The effect of
the radion is emphasized, and it is shown to disappear when the hidden brane is
taken to infinity, i.e. when we only have one brane. As a preliminary, we
derive the corresponding Lagrangian in the simpler geometry of a flat spacetime
with one extra dimension compactified on a torus. We also study the gauge
invariance of the theory, and note that the massless vector field that is
present for the torus disappears in the RS model because of the orbifold
symmetry.
\end{abstract}


\maketitle

\section{Introduction}

In this paper we develop the effective four-dimensional Lagrangian of gravity
in theories with extra dimensions, and in particular we study the fine-tuned
Randall-Sundrum (RS) model~\cite{RandallSundrum1, RandallSundrum2}. The
derivation include both the RS2 model with a single
brane~\cite{RandallSundrum2}, and also the more phenomenologically interesting
RS1 model with two branes~\cite{RandallSundrum1}. In the latter case, the
four-dimensional Lagrangian includes a massless scalar field (radion), a result
that is typical of theories with compact extra dimensions. The Lagrangian is
then used to calculate the Newtonian gravitational potential between two static
point sources on the visible brane.

Theories with extra dimensions, and in particular the braneworld model of
Randall and Sundrum, have attracted a lot of attention recently
\cite{Garriga_Giddings_Deruelle, Chung, Smolyakov,
Charmousis_Kiritsis_Nojiri_Ghoroku, Callin1, Callin2, Boos, Binetruy_Brevik}.
One central issue is whether effective four-dimensional gravity is restored at
the scales currently probed by experiments, i.e. whether the gravitational
potential $V(r) \approx -G m_1 m_2 / r$ at those scales. There seems to be
three different ways of calculating the potential. One way is to calculate the
component $h_{00}$ of the metric perturbation due to a matter source
directly~\cite{Garriga_Giddings_Deruelle, Chung, Smolyakov}. This method,
however, usually leads to complications with brane-bending, i.e. the fact that
the branes can no longer be considered straight when introducing a matter
source on them. A different approach is to consider the gravitational potential
as the result of two massive particles interacting through the exchange of a
virtual graviton, and to simply use the wave equation of the graviton to
determine its propagator and interaction with
matter~\cite{Charmousis_Kiritsis_Nojiri_Ghoroku, Callin1, Callin2}. With this
approach the branes are no longer bent, because the gravitons are actually
travelling through empty space. However, it can be difficult to fix the
normalization of the different four-dimensional fields from their wave
equations alone, and this normalization is crucial in order to get the correct
relative contribution from the radion as compared to the four-dimensional
gravitons. The most reliable and consistent approach is therefore to derive the
Lagrangian of the five-dimensional graviton, do a dimensional reduction to four
dimensions, and then identify the physical four-dimensional fields by requiring
that they have canonical Lagrangians~\cite{Boos}. We are going to use this
approach throughout the paper.

In section \ref{sec:Lagr_graviton} we derive the Lagrangians and propagators of
both massless and massive gravitons in $D$ spacetime dimensions, so that this
can be used as a reference to identify the physical graviton fields later on.
In section \ref{sec:torus} we study the simplest possible case with extra
dimensions, that is flat space with one dimension compactified on a torus. The
main features of the derivation will be the same as for the warped RS geometry,
only with much simpler algebra. This serves as a training ground before moving
on to the RS model in section~\ref{sec:RS}. Finally, we conclude in
section~\ref{sec:summary}.

\section{The Lagrangian and propagator of spin-2 fields in flat space}
\label{sec:Lagr_graviton}

\subsection{Massless spin-2 fields}
\label{sec:Lagr_massless}

With an arbitrary number $D$ of spacetime dimensions, the action of gravity is
given by
\be
  S = -\frac{1}{2} M^{D-2} \int d^D x \sqrt{g} R \, ,
  \label{eq:action_gravity}
\ee
where $M$ is the Planck mass in $D$ dimensions, and $R$ the scalar curvature.
The Lagrangian of gravitons in flat space is obtained by considering the
perturbation
\be
  g_{\mu\nu} = \eta_{\mu\nu} + 2 M^{(2-D)/2} h_{\mu\nu}
  \label{eq:pertmetric}
\ee
to the flat space metric~$\eta_{\mu\nu}$. The factor $M^{(2-D)/2}$ is chosen to
give $h_{\mu\nu}$ the correct physical dimension $\mathrm{dim}[h] =
\mbox{$(D-2)/2$}$, and the factor $2$ gives the canonical normalization of the
graviton field $h_{\mu\nu}$. Eq.~(\ref{eq:pertmetric}) is then inserted into
(\ref{eq:action_gravity}) and the result expanded to the second order
in~$h_{\mu\nu}$. After a straightforward calculation which involves several
partial integrations, we find $S = \int d^D x \, {\cal L}_h$, where the
graviton Lagrangian ${\cal L}_h$ is
\ba
  \hspace{-5mm} {\cal L}_h \! \eq \!
    - \tfrac{1}{2} h_{,\alpha} h^{,\alpha}
    \! + \! \tfrac{1}{2} h_{\alpha\beta,\nu} h^{\alpha\beta,\nu}
    \! + \! {h^{\alpha\nu}}_{,\alpha} h_{,\nu}
    \! - \! {h^{\alpha\nu}}_{,\alpha} {h^\beta}_{\nu,\beta} .
  \label{eq:graviton_Lagr}
\ea

So far, everything we have done has been valid in an arbitrary gauge. However,
in order to invert the quadratic operator and find the graviton propagator, we
must fix the gauge to be used in the Lagrangian. The Lagrangian
(\ref{eq:graviton_Lagr}) is invariant under the infinitesimal coordinate
transformation
\be
  x^\mu \to \bar{x}^\mu = x^\mu + 2 M^{(2-D)/2} \xi^\mu(x) \, ,
\ee
which transforms the graviton field as
\be
  h_{\mu\nu} \to \bar{h}_{\mu\nu} =
    h_{\mu\nu} - \left( \partial_\mu \xi_\nu + \partial_\nu \xi_\mu \right) .
\ee
The harmonic gauge where $\ned{\bar{h}}{\mu\nu}{,\nu} - \frac{1}{2}
\bar{h}^{,\mu} = 0$ is obtained by choosing $\Box \xi^\mu = {h^{\mu\nu}}_{,\nu}
- \frac{1}{2} h^{,\mu}$. In this gauge, we can therefore add the gauge fixing
term
\be
  {\cal L}_{hg} = \frac{1}{\xi} C_\mu C^\mu \, , \hspace{10mm}
  C^\mu = {h^{\mu\nu}}_{,\nu} - \frac{1}{2} h^{,\mu} \, ,
  \label{eq:gaugefix}
\ee
to the Lagrangian. The action then becomes
\ba
  S \eq \int d^D x \frac{1}{2} h^{\mu\nu} \left[
    \left(1-\frac{1}{2\xi}\right) \eta_{\mu\nu} \eta_{\alpha\beta} \partial^2
  \right. \nonumber \\
  && - \,
    \frac{1}{2} \left(
      \eta_{\mu\alpha} \eta_{\nu\beta} + \eta_{\mu\beta} \eta_{\nu\alpha}
    \right) \partial^2 \nonumber \\
  && + \,
    \left( \frac{1}{\xi} - 1 \right) \left(
      \eta_{\mu\nu} \partial_\alpha \partial_\beta +
      \eta_{\alpha\beta} \partial_\mu \partial_\nu
    \right) \nonumber \\
  && + \,
    \frac{1}{2} \left( 1 - \frac{1}{\xi} \right) \left( \addsmsp{
      \eta_{\mu\alpha} \partial_\nu \partial_\beta
      + \eta_{\mu\beta} \partial_\nu \partial_\alpha
    } \right. \nonumber \\
  && \left. \phantom{\frac{1}{2}} \hspace{18.8mm}
    \left. \addsmsp{
      + \, \eta_{\nu\alpha} \partial_\mu \partial_\beta
      + \eta_{\nu\beta} \partial_\mu \partial_\alpha
    } \right)
  \right] h^{\alpha\beta} \nonumber \\
  \eqequiv \int d^D x \frac{1}{2}
    h^{\mu\nu} \Box_{\mu\nu\alpha\beta} h^{\alpha\beta} \, .
\ea
The graviton propagator $D_{\alpha\beta\rho\sigma}(x,x')$ is defined as the
Green's function inverse to the operator $\Box_{\mu\nu\alpha\beta}$, i.e.
\be
  \Box_{\mu\nu\alpha\beta} D^{\alpha\beta\rho\sigma}(x,x') =
    \frac{1}{2} \left(
      \delta_\mu^{\;\rho} \delta_\nu^{\;\sigma} +
      \delta_\mu^{\;\sigma} \delta_\nu^{\;\rho}
    \right) \delta(x-x') \, .
  \label{eq:def_propagator}
\ee
The solution can then be written as
\be
  D_{\alpha\beta\rho\sigma}(x,x') = \int \frac{d^D k}{(2\pi)^D}
    \frac{P_{\alpha\beta\rho\sigma}(k)}{k^2} e^{-ik\cdot(x-x')} \, ,
\ee
where the polarization tensor $P_{\alpha\beta\rho\sigma}(k)$ is given by
\ba
  \hspace{-7mm} P_{\alpha\beta\rho\sigma}(k) \eq
    \frac{1}{2} \left(
      \eta_{\alpha\rho} \eta_{\beta\sigma} +
      \eta_{\alpha\sigma} \eta_{\beta\rho}
    \right) - \frac{1}{D-2} \eta_{\alpha\beta} \eta_{\rho\sigma} \nonumber \\
  && \hspace{-19.7mm} + \, \frac{(\xi-1)}{2 k^2} \! \left(
    \eta_{\alpha\rho} k_\beta k_\sigma \! + \!
    \eta_{\alpha\sigma} k_\beta k_\rho \! + \!
    \eta_{\beta\rho} k_\alpha k_\sigma \! + \!
    \eta_{\beta\sigma} k_\alpha k_\rho
  \right) .
  \label{eq:poltensor_massless}
\ea
Thus, we see that the de Donder gauge where $\xi = 1$ is a natural choice. The
result (\ref{eq:poltensor_massless}) is in agreement with~\cite{Giudice}.

The interaction between gravitons and matter which is described by the
Lagrangian ${\cal L}_m$, is obtained by including the action $S_m = \int d^D x
\sqrt{g} {\cal L}_m$. Varying this action with respect to the metric, we then
get $S_\mathrm{int} = \int d^D x \sqrt{g} \frac{1}{2} T_{\mu\nu} \delta
g^{\mu\nu}$, and thus
\be
  S_\mathrm{int} = -\int d^D x M^{(2-D)/2} T^{\mu\nu} h_{\mu\nu} \, ,
\ee
by using (\ref{eq:pertmetric}) to the lowest order in~$h_{\mu\nu}$.

\subsection{Massive spin-2 fields}
\label{sec:Lagr_massive}

The massless graviton in the previous section is given a mass by adding the
Fierz-Pauli term \mbox{$-\frac{1}{2} m^2 (h^{\alpha\beta} h_{\alpha\beta} \! -
\! h^2)$} to the Lagrangian~(\ref{eq:graviton_Lagr}). We thus have the
Lagrangian of a massive \mbox{spin-2} field
\ba
  {\cal L}^m_h \eq
    - \, \tfrac{1}{2} h_{,\alpha} h^{,\alpha}
    + \tfrac{1}{2} h_{\alpha\beta,\nu} h^{\alpha\beta,\nu}
    + {h^{\alpha\nu}}_{,\alpha} h_{,\nu} \nonumber \\
  \eqnl - \, {h^{\alpha\nu}}_{,\alpha} {h^\beta}_{\nu,\beta}
    - \tfrac{1}{2} m^2 \left( h^{\alpha\beta} h_{\alpha\beta} - h^2 \right) .
  \label{eq:graviton_Lagr_mass}
\ea
The resulting equation of motion for the free, massive graviton is
\ba
  && \hspace{-4mm} 0 = \Box^m_{\mu\nu\alpha\beta} h^{\alpha\beta} =
    \eta_{\mu\nu} (\partial^2 + m^2) h - (\partial^2 + m^2) h_{\mu\nu}
    \nonumber \\
  && - \, \eta_{\mu\nu} \partial_\alpha \partial_\beta h^{\alpha\beta}
    - \partial_\mu \partial_\nu h
    + \partial_\mu \partial^\alpha h_{\alpha\nu}
    + \partial_\nu \partial^\alpha h_{\alpha\mu} \, . \hspace{6mm}
\ea
Taking the derivative and contracting this equation, we obtain the constraints
\be
  \partial^\mu h_{\mu\nu} = 0 \, , \hspace{10mm}
  h = \eta^{\mu\nu} h_{\mu\nu} = 0 \, .
\ee
One thus obtains the simplified equation of motion
\be
  -\Box^m_{\mu\nu\alpha\beta} h^{\alpha\beta} =
    (\partial^2 + m^2) h_{\mu\nu} = 0 \, ,
\ee
which is why the Fierz-Pauli term seems a natural way of giving the graviton a
mass.

With this mass term included, there is no gauge freedom in the Lagrangian. The
operator $\Box^m_{\mu\nu\alpha\beta}$ can therefore be inverted directly,
leading to the propagator
\be
  D^m_{\alpha\beta\rho\sigma}(x,x') = \int \frac{d^D k}{(2\pi)^D}
    \frac{P^m_{\alpha\beta\rho\sigma}(k)}{k^2-m^2} e^{-ik\cdot(x-x')} \, ,
\ee
with the polarization tensor
\ba
  P^m_{\alpha\beta\rho\sigma}(k) \eq
    \tfrac{1}{2} \left(
      \eta_{\alpha\rho} \eta_{\beta\sigma} +
      \eta_{\alpha\sigma} \eta_{\beta\rho}
    \right) - \tfrac{1}{D-2} \eta_{\alpha\beta} \eta_{\rho\sigma} \nonumber \\
  && \hspace{-16mm} - \,
    \tfrac{1}{2m^2} \left(
      \eta_{\alpha\rho}k_\beta k_\sigma + \eta_{\alpha\sigma}k_\beta k_\rho +
      \eta_{\beta\rho} k_\alpha k_\sigma + \eta_{\beta\sigma} k_\alpha k_\rho
    \right) \nonumber \\
  && \hspace{-16mm} + \,
    \tfrac{1}{(D-1)(D-2)}
    \left( \eta_{\alpha\beta} + \tfrac{D-2}{m^2} k_\alpha k_\beta \right)
    \left( \eta_{\rho\sigma} + \tfrac{D-2}{m^2} k_\rho k_\sigma \right) .
    \nonumber \\
  && \label{eq:poltensor_massive}
\ea
This result was also found in \cite{Giudice} for the case $D=4$.

\section{Torus compactification}
\label{sec:torus}

We first consider the case of flat $(4+1)$-dimensional space, where the extra
dimension is compactified on a torus with circumference $L$. This allows us to
get comfortable with the formalism before moving on to the more complicated RS
model. It also allows us to do the whole calculation of the gravitational
potential from a purely five-dimensional point of view, so that we can compare
the results when doing a four-dimensional decomposition of the fields.
Corrections to the gravitational potential with torus compactification was also
studied in~\cite{Kehagias, Floratos_Liu}.

\subsection{Five-dimensional point of view}

From a five-dimensional point of view, the graviton is massless, and the
polarization tensor in the propagator is given by (\ref{eq:poltensor_massless})
with $D=5$. The interaction vertex is $M^{-3/2} T_{\mu\nu}$. For a point
particle with mass $m$ at rest at the point $(\boldx,y)$, $y$ being the
coordinate in the fifth dimension, the energy momentum tensor is
\be
  T_{\mu\nu}(\boldx,y) = m u_\mu u_\nu \delta(\boldx) \delta(y) =
    m \delta_\mu^0 \delta_\nu^0 \delta(\boldx) \delta(y) \, ,
\ee
and thus
\be
  T_{\mu\nu}(\boldk,p) = m \delta_\mu^0 \delta_\nu^0 \, ,
\ee
where $p$ is the momentum in the fifth dimension. The gravitational potential
between two point particles with masses $m_1$ and $m_2$ is therefore
\ba
  V(\boldk,p) \eq \lim_{k^0 \to 0}
    \!\! \raisebox{-9.5mm}{
         \includegraphics[bb= 270 722 352 781]{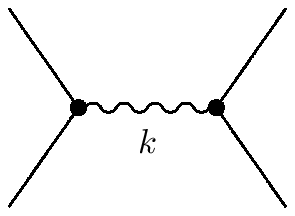}}
    \nonumber \\
  \eq \lim_{k^0 \to 0} \left[
    \frac{m_1 m_2}{M^3} \frac{P_{0000}(k)}{k_0^2 - \boldk^2 - p^2}
  \right] \nonumber \\
  \eq -\frac{2 m_1 m_2}{3 M^3} \frac{1}{\boldk^2 + p^2} \, .
  \label{eq:pot_torus_5d_k}
\ea
In particular, we note that the gauge dependent term in
(\ref{eq:poltensor_massless}) disappears, because only the $0000$-component of
the polarization tensor contributes, and $k^0 \to 0$. When Fourier inverting
(\ref{eq:pot_torus_5d_k}), the finite length of the fifth dimension means that
the momentum $p$ is quantized as $p_n = 2\pi n/L$, $n = 0, \pm 1, \pm 2,
\ldots$, with the result
\ba
  V(r,y) \eq \int \frac{d^3 k}{(2\pi)^3} \frac{1}{L} \sum_p
    V(\boldk,p) e^{i\boldk\cdot\boldx} e^{ipy} \nonumber \\
  \eq -\frac{m_1 m_2}{6\pi M^3 L r} \sum_p e^{-|p|r + ipy} \nonumber \\
  \eq -\frac{m_1 m_2}{6\pi M^3 L r}
    \frac{\sinh\frac{2\pi r}{L}}
         {\cosh\frac{2\pi r}{L}-\cos\frac{2\pi y}{L}} \, ,
  \label{eq:pot_torus_5d}
\ea
where $r = |\boldx|$. This is the same result as obtained
in~\cite{Floratos_Liu}.

Of particular importance is the limit of large distances, where the fifth
dimension disappears from the potential, and we are left with the
four-dimensional result
\ba
  \hspace{-6mm} \lim_{r \to \infty} \! V(r) \eq
    -\frac{m_1 m_2}{6\pi M^3 L r} \equiv
    -\frac{m_1 m_2}{8\pi \MPl^2 r} = -\frac{G m_1 m_2}{r} \, ,
\ea
where $\MPl$ is the effective four-dimensional Planck mass, and $G$ the
gravitational constant. We therefore obtain
\be
  \MPl^2 = \frac{3}{4} M^3 L \, .
  \label{eq:Planck4d_torus}
\ee
The factor $3/4$ may be a bit surprising. In the next section we will see that
it is the combination of a four-dimensional graviton and scalar field that both
contribute to the long range gravitational force that gives this factor. Of
course, it is possible to define the four-dimensional Planck mass by only
considering the \mbox{spin-2} fields that contribute to the long range force,
obtaining the common $\MPl^2 = M^3 L$. In that case, however, the potential
will no longer have the usual form $V(r) = -G m_1 m_2 / r$, but also an
additional contribution due to the scalar field. It therefore seems natural to
also include the scalar mode in the definition of the four-dimensional Planck
mass.

As a consistency check, we can also consider the limit of
(\ref{eq:pot_torus_5d}) for short distances, i.e. $r,y \to 0$:
\be
  \lim_{r,y \to 0} V(r,y) = -\frac{m_1 m_2}{6\pi^2 M^3 (r^2 + y^2)} \, .
\ee
This is precisely the potential in five-dimensional flat space, where all the
dimensions are large. From here we can also identify the five-dimensional
gravitational constant $G_5 = (3\pi^2 M^3)^{-1}$ \cite{footnote_potential}.

\subsection{Four-dimensional point of view}

We now decompose the five-dimensional graviton $h_{MN}$ from the previous
section into four-dimensional fields, and show that we obtain the same result
for the gravitational potential using only four-dimensional fields and
propagators. First the tensor structure is decomposed~as
\be
  h_{MN} = \left( \begin{matrix}
    h_{\mu\nu} & V_\mu \vs \\
    V_\nu & S
  \end{matrix} \right) .
  \label{eq:h_4dparts}
\ee
When inserted into (\ref{eq:graviton_Lagr}), we then obtain
\ba
  S \eq \int d^4 x \, dy \left\{ \addsmsp{
    - \, V_{\alpha,\beta} V^{\alpha,\beta}
    + V_{\alpha,\beta} V^{\beta,\alpha}
  } \right. \nonumber \\
  \eqnl
    - \, \tfrac{1}{2} h_{,\alpha} h^{,\alpha}
    + \tfrac{1}{2} h_{\alpha\beta,\nu} h^{\alpha\beta,\nu}
    + {h^{\alpha\nu}}_{,\alpha} h_{,\nu}
    - {h^{\alpha\nu}}_{,\alpha} {h^\beta}_{\nu,\beta} \nonumber \\
  \eqnl \hspace{-0.4mm} \left. \addsmsp{
    - \, \tfrac{1}{2} \left[ h'_{\alpha\beta} h'^{\alpha\beta} - (h')^2 \right]
    + S_{,\alpha} h^{,\alpha} - {h^{\alpha\beta}}_{,\beta} S_{,\alpha}
  } \right\} ,
  \label{eq:action_torus_cross}
\ea
where the prime means the derivative with respect to~$y$. In particular, we
note that all terms with $y$-derivatives of $V^\mu$ and $S$ have cancelled,
which means that only the graviton $h_{\mu\nu}$ will acquire a mass in four
dimensions.

This is related to the gauge freedom of the five-dimensional theory, where a
possible gauge is to demand that both $V^\mu$ and $S$ are independent of $y$,
as we will now show. Starting with the five-dimensional coordinate
transformation
\be
  x^M \to \bar{x}^M = x^M + 2 M^{-3/2} \xi^M \, ,
\ee
the five-dimensional graviton field is transformed as
\be
  h_{MN} \to \bar{h}_{MN} =
    h_{MN} - \left( \partial_M \xi_N + \partial_N \xi_M \right) .
\ee
Using (\ref{eq:h_4dparts}), we then get for the different components
\ba
  \bar{h}_{\mu\nu} \eq h_{\mu\nu} -
    \left( \partial_\mu \xi_\nu + \partial_\nu \xi_\mu \right) , \nonumber \\
  \bar{V}_\mu \eq V_\mu - \partial_\mu \xi_4 - \xi'_\mu \, , \nonumber \\
  \bar{S} \eq S - 2 \xi'_4 \, .
\ea
By choosing a suitable function $\xi_4$ we can simplify the radion~$S$. It is
not possible to gauge the field away since $\xi_4$ must be periodic under $y
\to L+y$, but the choice
\ba
  \xi_4(x,y) \eq \frac{1}{2} \int_0^y \! S(x,y') dy' -
    \frac{y}{2L} \int_0^L \! S(x,y') dy' + \bar{\xi}(x) \, , \nonumber \\
  && \label{eq:xi4_torus}
\ea
with $\bar{\xi}(x)$ arbitrary, removes the $y$-dependence of~$S$:
\be
  \bar{S}(x,y) = \frac{1}{L} \int_0^L \! S(x,y') dy' = \bar{S}(x) \, .
\ee
We also see that $\xi_4$ clearly satisfies the correct periodic boundary
conditions. Hence, by a choice of coordinates for the extra dimension, the
radion can be taken to be independent of~$y$. This is actually quite natural,
since the physical interpretation of the radion is that it measures the
physical size of the torus -- a quantity that necessarily depends only on the
four-dimensional coordinate~$x$.

Similarly, we choose $\xi_\mu$ to remove the $y$-dependence from the vector
field~$V_\mu$. Again, it is not possible to remove the vector field completely
since $\xi_\mu$ must be periodic. With the notation $W_\mu \equiv V_\mu -
\partial_\mu \xi_4$, we write
\ba
  \xi_\mu(x,y) \eq \! \int_0^y \!\! W_\mu(x,y') dy' -
    \frac{y}{L} \! \int_0^L \!\! W_\mu(x,y') dy' + \bar{\xi}_\mu(x) \, ,
    \nonumber \\
  && \label{eq:ximu_torus}
\ea
with $\bar{\xi}_\mu(x)$ arbitrary. The transformed vector field is then
\be
  \bar{V}_\mu(x,y) =
    \frac{1}{L} \int_0^L \! W_\mu(x,y') dy' = \bar{V}_\mu(x) \, .
\ee
A choice of four-dimensional coordinates can therefore make the vector field
independent of~$y$. We call the gauge where both $V_\mu$ and $S$ are
independent of $y$ for the \textit{unitary gauge}, and use it in the following.
Note that the zero modes of both $V^\mu$ and $S$ are physical degrees of
freedom with torus compactification. It is not possible to choose a gauge where
either of them disappears completely.

The graviton field $h_{\mu\nu}$ is expanded in plane waves:
\be
  h_{\mu\nu}(x,y) = \sum_p h^p_{\mu\nu}(x) \psi_p(y) \, , \hspace{3mm}
  \psi_p(y) = \frac{1}{\sqrt{L}} e^{ipy} \, .
  \label{eq:graviton_torus}
\ee
In order to remove the crossterms between $h_{\mu\nu}$ and $S$ in
(\ref{eq:action_torus_cross}), which only affects the zero mode $h^0_{\mu\nu}$
since $S$ is independent of $y$, we redefine $h^0_{\mu\nu}$ as
\be
  h^0_{\mu\nu} \to h^0_{\mu\nu} + \frac{1}{\sqrt{6}} \eta_{\mu\nu} \phi \, ,
  \label{eq:graviton_torus_redef}
\ee
where we have also rescaled the scalar field as
\be
  S = \sqrt{\frac{2}{3L}} \, \phi \, ,
\ee
in order to get a canonical Lagrangian for~$\phi$. If we also make a trivial
rescaling $V^\mu = A^\mu / \sqrt{2L}$ of the vector field, the action is
reduced to
\ba
  S \eq \int d^4 x \left\{ \addsmsp{
    - \, \tfrac{1}{2} A_{\mu,\nu} A^{\mu,\nu}
    + \tfrac{1}{2} A_{\mu,\nu} A^{\nu,\mu}
    + \tfrac{1}{2} \phi_{,\mu} \phi^{,\mu}
  } \right. \nonumber \\
  \eqpluss \raisebox{0mm}[4mm][2.5mm]{$\ds \! \sum_p \! \left[
    - \, \tfrac{1}{2} h^p_{,\alpha} h_{-p}^{,\alpha}
    + \tfrac{1}{2} h^p_{\alpha\beta,\nu} h_{-p}^{\alpha\beta,\nu}
    + h^{\alpha\nu}_{p\;\;,\alpha} h^{-p}_{,\nu}
  \right. $} \nonumber \\
  \eqnl \hspace{5.3mm} \left. \addsmsp{ \left.
    - \, h^{\alpha\nu}_{p\;\;,\alpha} h^\beta_{-p\,\nu,\beta}
    - \tfrac{1}{2} p^2 \! \left(
      h^p_{\alpha\beta} h_{-p}^{\alpha\beta} \! - \! h^p h^{-p}
    \right)
  \right] } \right\} \! , \hspace{7mm}
  \label{eq:action_torus_eff}
\ea
after performing the $y$-integration. From this effective four-dimensional
action we can clearly identify the physical fields as a massless graviton
($p=0$), a tower of massive gravitons with Lagrangians identical to
(\ref{eq:graviton_Lagr_mass}), a massless vector field, and a massless scalar
field.

As we can see from (\ref{eq:xi4_torus}) and (\ref{eq:ximu_torus}), the unitary
gauge is not uniquely determined -- we can still make transformations in the
form of the two functions $\bar{\xi}(x)$ and $\bar{\xi}_\mu(x)$ which only
depend on the four-dimensional coordinate~$x$. This corresponds to the
transformations
\ba
  h^0_{\mu\nu} \eqto h^0_{\mu\nu} -
    \left( \partial_\mu \bar{\xi}_\nu + \partial_\nu \bar{\xi}_\mu \right) ,
    \nonumber \\
  A_\mu \eqto A_\mu - \partial_\mu \bar{\xi} \, ,
\ea
of the zero mode of the graviton and the vector field, and is precisely what
allows us to identify these fields as massless graviton and vector fields in
four dimensions. Also note that both actions (\ref{eq:action_torus_cross}) and
(\ref{eq:action_torus_eff}) are invariant under this residual gauge
transformation, meaning that the order which we perform the gauge
transformation and the field redefinition (\ref{eq:graviton_torus_redef})
doesn't matter.

The interaction with matter in five dimensions can be written $S_\mathrm{int} =
-\int d^4 x \, dy M^{-3/2} h_{MN} T^{MN}$. For a source located at $y=y'$ the
energy momentum tensor is
\be
  T^{MN}(x,y) = \delta^M_\mu \delta^N_\nu T^{\mu\nu}(x) \delta(y-y') \, .
  \label{eq:source_y}
\ee
Using the expansion (\ref{eq:graviton_torus}) and the redefinition
(\ref{eq:graviton_torus_redef}) we therefore get the effective four-dimensional
interaction
\be
  {\cal L}_\mathrm{int} = -\frac{M^{-3/2}}{\sqrt{L}} \left(
    \sum_p e^{ipy'} h^p_{\mu\nu} T^{\mu\nu} + \frac{1}{\sqrt{6}} \phi T
  \right) .
  \label{eq:int_torus}
\ee
The radion field thus couples to the trace of the energy momentum tensor.

From (\ref{eq:action_torus_eff}) and (\ref{eq:int_torus}) we can now obtain the
gravitational potential in the effective four-dimensional theory:
\ba
  V(\boldk,y) \eq
    \!\!\! \raisebox{-5.7mm}{
         \includegraphics[bb= 285 742 337 779]{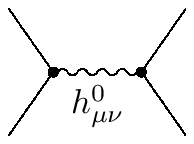}} +
    \sum_{p \neq 0} \!\!\! \raisebox{-5.7mm}{
         \includegraphics[bb= 285 742 337 779]{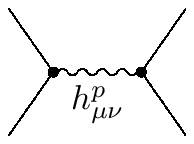}} +
    \!\!\! \raisebox{-5.7mm}{
         \includegraphics[bb= 285 742 337 779]{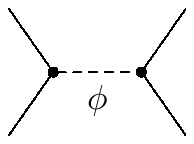}} \;
    \nonumber \\
  \eq -\frac{m_1 m_2}{M^3 L} \left[
    \frac{1}{2} \frac{1}{\boldk^2} +
    \frac{2}{3} \sum_{p \neq 0} \frac{e^{ipy}}{\boldk^2 + p^2} +
    \frac{1}{6} \frac{1}{\boldk^2}
  \right] \nonumber \\
  \eq -\frac{2 m_1 m_2}{3 M^3 L}
    \sum_{\mathrm{all} \; p} \frac{e^{ipy}}{\boldk^2+p^2} \, .
  \label{eq:pot_torus_4d}
\ea
The factors $1/2$ and $2/3$ for the gravitons originate from the polarization
tensors (\ref{eq:poltensor_massless}) and (\ref{eq:poltensor_massive}) for
$D=4$, whereas the factor $1/6$ for the radion follows from the
interaction~(\ref{eq:int_torus}). Thus, we see that the massless graviton and
the radion combined give exactly the same overall factor as the massive
gravitons. By Fourier inverting (\ref{eq:pot_torus_4d}), we get the same result
as (\ref{eq:pot_torus_5d}) derived in the previous section from the
five-dimensional point of view.

\section{The Randall-Sundrum model with two branes}
\label{sec:RS}

In the fine-tuned or critical RS model, the background metric is given by
\ba
  ds^2 \eq A^2(y) \eta_{\mu\nu} dx^\mu dx^\nu - dy^2 \nonumber \\
  \eq A^2(z) \left( \eta_{\mu\nu} dx^\mu dx^\nu - dz^2 \right) ,
  \label{eq:RS_backgrmetric}
\ea
where
\be
  A(y) = e^{-\mu|y|} \, , \hspace{5mm}
  A(z) = \frac{1}{1 + \mu|z|} \, ,
  \label{eq:RS_A}
\ee
is the warp factor in the physical coordinate $y$ and the conformal coordinate
$z$, respectively. The visible brane is located at $y=z=0$ and the hidden brane
at $y=y_r$, $z=z_r$. We assume the usual orbifold symmetry $S^1 / Z_2$ for the
fifth dimension, which basically means that the branes are located at the
endpoints of this dimension. The visible and hidden brane have tension $\lambda
= 6\mu M^3$ and $-\lambda$, respectively, and the cosmological constant in the
anti de Sitter bulk space is $\Lambda_B = -6\mu^2$. The fine-tuning between the
positive tension on the visible brane and the negative cosmological constant in
the bulk means that the effective four-dimensional cosmological constant
vanishes, and the four-dimensional space is therefore flat with
metric~$\eta_{\mu\nu}$. For a more detailed discussion on the phenomenology of
the RS model, see e.g.~\cite{Binetruy_Brevik}.

\subsection{Graviton Lagrangian}

Using the conformal $z$-coordinate, we write the perturbed metric as
\be
  g_{MN} = A^2(z) \left( \eta_{MN} + h_{MN} \right) .
\ee
The derivation of the Lagrangian for $h_{MN}$ is easiest to do using this
coordinate, but still involves some rather tedious calculations (see appendix
\ref{sec:conform_Lagr}). The starting point is the action
\ba
  S \eq -\frac{1}{2} M^3 \int d^4 x \, dz
    \mbox{$\sqrt{|g_{MN}|}$} \left( R + 2\Lambda_B \right) \nonumber \\
  \eqnl -\lambda \int d^4 x \, dz \mbox{$\sqrt{|g_{\mu\nu}|}$}
    \left[ \addsmsp{\delta(z) - \delta(z-z_r)} \right] .
  \label{eq:RS_action_general}
\ea
Here $\sqrt{|g_{MN}|}$ means the square root of the determinant of the
five-dimensional metric, i.e.
\ba
  \mbox{$\sqrt{|g_{MN}|}$} \eq A^5 \left[ \addsmsp{
    1 + \tfrac{1}{2} (h-S) + \tfrac{1}{8} (h-S)^2
  } \right. \nonumber \\
  && \hspace{4.9mm} \left. \addsmsp{
    - \, \tfrac{1}{4} h^{\mu\nu} h_{\mu\nu}
    + \tfrac{1}{2} V^\mu V_\mu - \tfrac{1}{4} S^2
  } \right] ,
\ea
where we have used (\ref{eq:h_4dparts}) for $h_{MN}$, whereas
$\sqrt{|g_{\mu\nu}|}$ only involves the four-dimensional components, i.e.
\be
  \mbox{$\sqrt{|g_{\mu\nu}|}$} = A^4 \left[ \addsmsp{
    1 + \tfrac{1}{2} h + \tfrac{1}{8} h^2 - \tfrac{1}{4} h^{\mu\nu} h_{\mu\nu}
  } \right] .
\ee

From the background solution (\ref{eq:RS_A}) it follows that both the zeroth
order and the first order parts of (\ref{eq:RS_action_general}) vanish, as they
should. For the second order part, the result (\ref{eq:action_general}) of
appendix \ref{sec:conform_Lagr} then implies
\ba
  S \eq -\frac{1}{2} M^3 \int d^4 x \, dy \left\{ \addsmsp{
    A^2 \! \left[ \addsmsp{
      \tfrac{1}{2} V_{\alpha,\beta} V^{\alpha,\beta}
      - \tfrac{1}{2} V_{\alpha,\beta} V^{\beta,\alpha}
    } \right]
  } \right. \nonumber \\
  && \hspace{-8mm} + \, A^2 \! \left[ \addsmsp{
    - \, \tfrac{1}{4} h_{\alpha\beta,\nu} h^{\alpha\beta,\nu}
    - \tfrac{1}{2} {h^{\alpha\beta}}_{,\beta} h_{,\alpha}
    + \tfrac{1}{2} {h^{\alpha\nu}}_{,\alpha} {h^\beta}_{\nu,\beta}
  } \right. \nonumber \\
  && \hspace{0.9mm} \left. \addsmsp{
    + \, \tfrac{1}{4} h_{,\alpha} h^{,\alpha}
    + \tfrac{1}{2} {h^{\alpha\beta}}_{,\beta} S_{,\alpha}
    - \tfrac{1}{2} h_{,\alpha} S^{,\alpha}
  } \right] \nonumber \\
  && \hspace{-8.4mm} \left. \addsmsp{
    + \, A^4 \! \left[ \addsmsp{
      \tfrac{1}{4} h'_{\alpha\beta} h'^{\alpha\beta}
      \! - \! \tfrac{1}{4} (h')^2
    } \right]
    - \tfrac{3}{2} A^3 A' S h' - 3\mu^2 A^4 S^2
  } \right\} \! , \hspace{7mm}
  \label{eq:RS_action}
\ea
where we have also changed the integration variable to $y$, and the prime means
derivative with respect to~$y$. Again we note that all terms with
$y$-derivatives of $V^\mu$ and $S$ have disappeared, so the graviton field will
be the only massive field in four dimensions (the $S^2$-term will be removed by
a field redefinition). However, in contrast to the torus, it is now possible to
choose a gauge where the vector field disappears completely.

The issue of gauge transformations in the RS model has been discussed in detail
by Boos \textit{et al.} \cite{Boos}, but we include the main parts here for
completeness. In the RS braneworld there is an additional symmetry that is not
present for the torus, namely the orbifold symmetry $S^1 / Z_2$. As we will
see, this symmetry is what removes the vector field as a physical degree of
freedom. This conclusion can actually be reached almost trivially by just
looking at the orbifold symmetries that the components of the five-dimensional
graviton must satisfy:
\ba
  h_{\mu\nu}(x,-z) \eq h_{\mu\nu}(x,z) \, , \nonumber \\
  V_\mu(x,-z) \eq -V_\mu(x,z) \, , \nonumber \\
  S(x,-z) \eq S(x,z) \, .
  \label{eq:orbifold_fields}
\ea
We expect that the $z$-dependence of the vector field can be removed just as
for the torus, but additionally, since $V_\mu(z)$ is an odd function of $z$, it
cannot have a zero mode either. Thus, it should be possible to choose a gauge
where the vector field disappears altogether.

To see this in more detail, we start with the five-dimensional coordinate
transformation $x^M \to \bar{x}^M = x^M + \xi^M$, which transforms the graviton
as
\be
  h_{MN} \to \bar{h}_{MN} = h_{MN} -
    A^{-2} \! \left( \nabla_M \xi_N + \nabla_N \xi_M \right) ,
\ee
where the covariant derivative is with respect to the background metric $g_{MN}
= A^2(z) \eta_{MN}$. Writing the gauge functions as
\be
  \xi_\mu = A^2 \hat{\xi}_\mu \, , \hspace{10mm}
  \xi_4 = A \hat{\xi}_4 \, ,
\ee
for convenience, the components of $h_{MN}$ transform as
\ba
  \bar{h}_{\mu\nu} \eq h_{\mu\nu} -
    \left(
      \partial_\mu \hat{\xi}_\nu + \partial_\nu \hat{\xi}_\mu -
      \frac{2A'}{A^2} \eta_{\mu\nu} \hat{\xi}_4
    \right) , \nonumber \\
  \bar{V}_\mu \eq V_\mu -
    \left( \frac{1}{A} \partial_\mu \hat{\xi}_4 + \hat{\xi}'_\mu \right) ,
    \nonumber \\
  \bar{S} \eq S - \frac{2}{A} \hat{\xi}'_4 \, ,
\ea
where the prime means the derivative with respect to~$z$. We choose the gauge
where
\be
  \hat{\xi}_\mu(x,-z) = \hat{\xi}_\mu(x,z) \, , \hspace{5mm}
  \hat{\xi}_4(x,-z) = -\hat{\xi}_4(x,z) \, .
\ee
The $z$-dependence of the radion is removed by choosing
\ba
  \hat{\xi}_4(x,z) \eq \frac{1}{2} \! \int_0^z \!\!\! A(z') S(x,z') dz' -
    \frac{y}{2 y_r} \! \int_0^{z_r} \!\!\!\!\! A(z') S(x,z') dz' .
    \nonumber \\
  && \label{eq:xi4_RS}
\ea
We then get, by using $dy/dz = A(z)$,
\ba
  \bar{S}(x,z) \eq \frac{1}{y_r} \int_0^{z_r} \!\! A(z') S(x,z') dz'
    \nonumber \\
  \eq \frac{1}{y_r} \int_0^{y_r} \!\! S(x,y') dy' = \bar{S}(x) \, .
\ea
Note that we can not add an arbitrary function $\bar{\xi}(x)$ in
(\ref{eq:xi4_RS}), in contrast to the torus, since this would break the
orbifold symmetry of~$\hat{\xi}_4$. This is another hint that the vector field
must disappear, since we don't have any residual gauge transformations left
that would transform the massless four-dimensional vector field. Also note that
we impose the orbifold symmetry both before and after the coordinate
transformation, which means that the visible brane remains straight in position
$\bar{z} = 0$. This is also clearly seen from the fact that $\hat{\xi}_4(z=0) =
0$. Thus, we do not consider brane bending here, since brane bending can always
be removed by a coordinate transformation~\cite{Garriga_Giddings_Deruelle}.

Moving on to the vector field, we see that we obtain $\bar{V}_\mu = 0$ if we
choose
\be
  \hat{\xi}_\mu(x,z) = \int_0^z \! W_\mu(x,z') dz' + \bar{\xi}_\mu(x) \, ,
  \label{eq:ximu_RS}
\ee
where $W_\mu \equiv V_\mu - A^{-1} \partial_\mu \hat{\xi}_4$, and the function
$\bar{\xi}_\mu(x)$ can be chosen arbitrarily. The reason why this works in the
RS model but not for the torus, is that the function $W_\mu$ is now an odd
function of $z$, which implies both the correct orbifold symmetry
$\hat{\xi}_\mu(-z) = \hat{\xi}_\mu(z)$, and the periodic boundary condition
$\hat{\xi}_\mu(2z_r+z) = \hat{\xi}_\mu(z)$. The latter follows from the
identity $\int_0^{2z_r} \! W_\mu(z) dz = \int_{-z_r}^{z_r} \! W_\mu(z) dz = 0$.
Thus, we see that the vector field is a pure gauge mode in the RS model, and by
choosing the four-dimensional coordinates favourably we can make the field
vanish. We call the gauge where $V_\mu = 0$ and $S$ is independent of $z$ (or
$y$) for the \textit{unitary gauge} in the RS model, and use it in the
following.

Next, we turn to the crossterms between $h_{\mu\nu}$ and~$S$. Because they are
a bit more involved than for the torus, we get the more complicated
redefinition
\be
  h_{\mu\nu} \to h_{\mu\nu} + f(y) \eta_{\mu\nu} S +
    g(y) \partial_\mu \partial_\nu S \, ,
  \label{eq:RS_h_redef}
\ee
where
\ba
  f(y) \eq \mu|y| + f_0 \, , \hspace{10mm}
    f_0 = \frac{\mu y_r}{e^{2\mu y_r} - 1} \, , \label{eq:RS_fg} \\
  g(y) \eq \frac{1}{4\mu^2} \! \left\{ \!
    f_0 \! \left[ e^{2\mu|y|} \! - \! 1 \right]^2 \!\! +
    e^{2\mu|y|} \! \left[ 1 \! - \! 2\mu|y| \right] - 1
  \! \right\} \! + g_0 \, . \nonumber
\ea
The constant $f_0 \equiv f(0)$ is determined by requiring that the crossterms
disappear on the branes, i.e. that we don't have any delta function crossterms.
(More precisely, this implies the conditions $g'(0^+) = g'(y_r^-) = 0$.) The
constant $g_0 \equiv g(0)$, however, doesn't seem to have any constraints, and
can therefore be chosen freely. In any case, its value will not affect the
gravitational potential.

With the field redefinition (\ref{eq:RS_h_redef}), all crossterms between
$h_{\mu\nu}$ and $S$ and also the $S^2$ term disappear from
(\ref{eq:RS_action}), and we are left with a kinetic term $S_{,\alpha}
S^{,\alpha}$ for the scalar field with a coefficient
\be
  -\frac{1}{2} M^3 \int_{-y_r}^{y_r} \frac{3}{2} \left( f^2 - f \right) A^2 dy
  = \frac{\tfrac{3}{4} M^3 \mu y_r^2}{e^{2\mu y_r} - 1} \, .
\ee
The physical radion $\phi$ with a kinetic term $\frac{1}{2} \phi_{,\alpha}
\phi^{,\alpha}$ must therefore be given by
\be
  \phi = \sqrt{\frac{\tfrac{3}{2} M^3 \mu y_r^2}{e^{2\mu y_r}-1}} \, S \, .
  \label{eq:RS_radion}
\ee
In particular, we note that the radion drops out of the action in the limit
$y_r \to \infty$ of one brane, as also shown in~\cite{Boos}. Finally we expand
$h_{\mu\nu}$ in terms of four-dimensional eigenstates
\be
  h_{\mu\nu}(x,y) = 2 M^{-3/2} \sum_m h^m_{\mu\nu}(x) \Phi_m(y) \, ,
  \label{eq:RS_graviton}
\ee
where the functions $\Phi_m(y)$ satisfy the following wave equation and
normalization:
\ba
  \Phi''_m(y) + \frac{4A'}{A} \Phi'_m(y) + \frac{m^2}{A^2} \Phi(y) \eq 0 \, ,
    \nonumber \\
  \int_{-y_r}^{y_r} A^2(y) \Phi_m(y) \Phi_{m'}(y) dy \eq \delta_{m,m'} \, ,
    \nonumber \\
  \int_{-y_r}^{y_r} A^4(y) \Phi'_m(y) \Phi'_{m'}(y) dy \eq m^2 \delta_{m,m'}
    \, . \hspace{3mm}
  \label{eq:RS_eigenfunc}
\ea
(The last of these equations follows trivially from the first two.) The general
solution to (\ref{eq:RS_eigenfunc}) can be expressed in terms of Bessel
functions, see e.g. \cite{Callin1} for more details. Combining all of this, we
are finally able to reduce (\ref{eq:RS_action}) to the effective
four-dimensional action
\ba
  S \eq \! \int \! d^4 x \left\{
    \addspace{ \sum_m \left[ \addsmsp{
      - \, \tfrac{1}{2} h^m_{,\alpha} h_m^{,\alpha}
      + \tfrac{1}{2} h^m_{\alpha\beta,\nu} h_m^{\alpha\beta,\nu}
      + h^{\alpha\nu}_{m\;,\alpha} h^m_{,\nu}
    } \right. }
  \right. \nonumber \\
  && \hspace{17.35mm} \raisebox{0mm}[2.5mm][1.5mm]{$\ds \left. \addsmsp{
    - \, h^{\alpha\nu}_{m\;,\alpha} h^\beta_{m\,\nu,\beta}
    - \tfrac{1}{2} m^2 \left(
      h^m_{\alpha\beta} h_m^{\alpha\beta} - h_m^2
    \right)
  } \right] $} \nonumber \\
  && \hspace{9.9mm} \left. \addspace{
    + \, \tfrac{1}{2} \phi_{,\mu} \phi^{,\mu}
  } \right\} \! .
  \label{eq:RS_action_eff}
\ea
Again we recognize this as a massless graviton \mbox{($m=0$)}, a tower of
massive gravitons, and a massless scalar field. The four-dimensional field
content of the theory is therefore the same as with torus compactification,
except for the vector field which is now absent.

From (\ref{eq:xi4_RS}) and (\ref{eq:ximu_RS}) we note that the only residual
gauge symmetry we have left after imposing the unitary gauge, is a
transformation of the four-dimensional coordinate by a function
$\bar{\xi}_\mu(x)$ which only depends on~$x$. The graviton field is then
transformed as
\be
  h^0_{\mu\nu} \to h^0_{\mu\nu} -
    \left( \partial_\mu \bar{\xi}_\nu + \partial_\nu \bar{\xi}_\mu \right) .
\ee
This is precisely the transformation that allows us to identify the zero mode
$h^0_{\mu\nu}$ of the five-dimensional graviton as a massless graviton in four
dimensions. Again we also note that both actions (\ref{eq:RS_action}) and
(\ref{eq:RS_action_eff}) are invariant under this residual gauge
transformation, so we can choose freely the order which we perform the field
redefinition (\ref{eq:RS_h_redef}) and the gauge transformation.

\subsection{Interaction with matter}

With the source (\ref{eq:source_y}), the interaction is given by
\ba
  S_\mathrm{int} \eq \int d^4 x \, dy \sqrt{g} \frac{1}{2} T_{MN} \delta g^{MN}
    \nonumber \\
  \eq -\int d^4 x \left[
    \frac{1}{2} A^2(y') T^{\mu\nu}(x) h_{\mu\nu}(x,y')
  \right] .
\ea
From (\ref{eq:RS_h_redef}), (\ref{eq:RS_radion}) and (\ref{eq:RS_graviton}), we
have
\ba
  h_{\mu\nu}(x,y') \eq 2 M^{-3/2} \sum_m h^m_{\mu\nu}(x) \Phi_m(y') +
    \nonumber \\
  && \hspace{-14mm}
    \sqrt{\frac{e^{2\mu y_r}-1}{\tfrac{3}{2} M^3 \mu y_r^2}} \left[
      \addsmsp{f(y') \eta_{\mu\nu} + g(y') \partial_\mu \partial_\nu}
    \right] \! \phi(x) \, . \hspace{5mm}
\ea
Since the energy momentum tensor $T_{\mu\nu}$ is conserved, $\partial_\mu
T^{\mu\nu} = 0$, the term with $g(y')$ disappears after a partial integration,
and we are left with
\ba
  {\cal L}_\mathrm{int} \eq -M^{-3/2} A^2(y') \nonumber \\
  && \hspace{-8mm} \times \! \left[
    \sum_m \Phi_m(y') h^m_{\mu\nu} T^{\mu\nu} +
    f(y') \sqrt{\frac{e^{2\mu y_r}-1}{6\mu y_r^2}} \, \phi T
  \right] \! . \hspace{6mm}
  \label{eq:RS_interaction}
\ea
Again we see that the radion couples to the trace of the energy momentum
tensor.

\subsection{The gravitational potential}

With one source on the brane at $y=0$ and the other at the point $y$, the
gravitational potential is
\ba
  V(\boldk,y) \eq
    \!\!\! \raisebox{-5.7mm}{
         \includegraphics[bb= 285 742 337 779]{Newton_Lagr_Feyn2.ps}} +
    \sum_{m>0} \!\!\! \raisebox{-5.7mm}{
         \includegraphics[bb= 285 742 337 779]{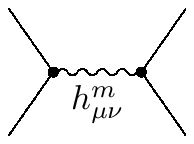}} +
    \!\!\! \raisebox{-5.7mm}{
         \includegraphics[bb= 285 742 337 779]{Newton_Lagr_Feyn4.ps}} \;
    \nonumber \\
  \eq -\frac{m_1 m_2}{M^3 e^{2\mu y}} \! \left\{ \!
    \frac{\Phi_0(0) \Phi_0(y)}{2\boldk^2} +
    \frac{2}{3} \!\sum_{m>0}\!\! \frac{\Phi_m(0) \Phi_m(y)}{\boldk^2 + m^2}
  \right. \nonumber \\
  && \hspace{10.6mm} \left. \phantom{\sum_m}
    + \frac{e^{2\mu y_r} - 1}{6\mu y_r^2} \frac{f(0) f(y)}{\boldk^2}
  \! \right\} .
\ea
With $f(y)$ from (\ref{eq:RS_fg}) this implies
\ba
  V(r,y) \eq -\frac{m_1 m_2}{8\pi M^3 e^{2\mu y} \, r} \left\{
    \frac{\mu}{1 - e^{-2\mu y_r}} + \phantom{\sum_m}
  \right. \nonumber \\
  && \hspace{-15mm} \left.
    \frac{4}{3} \!\sum_{m>0}\! \Phi_m(0) \Phi_m(y) e^{-mr} +
    \frac{1}{3} \! \left( \!
      \frac{\mu y}{y_r} + \frac{\mu e^{-2\mu y_r}}{1 - e^{-2\mu y_r}}
    \! \right) \!
  \right\} \! . \hspace{7mm}
  \label{eq:RS_pot_y}
\ea
Here we have inserted the expression
\be
  \Phi_0(y) = \sqrt{\frac{\mu}{1 - e^{-2\mu y_r}}}
\ee
for the zero mode, which follows from~(\ref{eq:RS_eigenfunc}). With both
sources on the visible brane, i.e. by setting $y=0$, (\ref{eq:RS_pot_y}) is
reduced to
\ba
  V(r) \eq -\frac{m_1 m_2}{8\pi M^3 r} \left\{
    \frac{\mu}{1 - e^{-2\mu y_r}} +
    \frac{4}{3} \sum_{m>0} \Phi_m^2(0) e^{-mr}
  \right. \nonumber \\
  && \hspace{10.9mm} \left. \phantom{\sum_m}
    + \, \frac{1}{3} \frac{\mu e^{-2\mu y_r}}{1 - e^{-2\mu y_r}}
  \right\} .
  \label{eq:RS_pot1}
\ea
Both the massless graviton and the radion contribute to the long range
potential. The radion contribution is, however, suppressed by an extra factor
$e^{-2\mu y_r} \sim (\MEW / \MPl)^2 \sim 10^{-30}$ if we interpret the RS1
model as a possible solution to the hierarchy problem. If we include the radion
in the definition of the effective four-dimensional Planck mass, we get
\be
  \MPl^2 = \frac{M^3}{\mu}
    \frac{1 - e^{-2\mu y_r}}{1 + \frac{1}{3} e^{-2\mu y_r}} \, ,
  \label{eq:RS_Planckmass}
\ee
and (\ref{eq:RS_pot1}) can then be written
\ba
  V(r) \eq -\frac{G m_1 m_2}{r} \! \left\{ \! 1 + \frac{4}{3\mu}
    \frac{1 - e^{-2\mu y_r}}{1 \! + \! \frac{1}{3} e^{-2\mu y_r}}
      \!\sum_{m>0}\! \Phi_m^2(0) e^{-mr}
  \! \right\} \! , \nonumber \\
  && \label{eq:RS_pot2}
\ea
where $G = (8\pi\MPl^2)^{-1}$ is the gravitational constant. The whole effect
of the radion is the factor $\left( 1 + \frac{1}{3} e^{-2\mu y_r} \right)$ in
the potential and in the four-dimensional Planck mass. The same factor also
appears by the related calculation of similar effects in~\cite{Smolyakov}.
Again we see that the radion contribution disappears in the limit $y_r \to
\infty$. In this limit, we can also show that~\cite{Callin1}
\be
  \lim_{y_r \to \infty} \Phi_m^2(0) = \frac{2\mu}{\pi m z_r}
    \frac{1}{J_1^2(\frac{m}{\mu}) + Y_1^2(\frac{m}{\mu})} \, ,
\ee
whereas $\sum_m \to \int_0^\infty \frac{z_r}{\pi} dm$, with the
result~\cite{Chung, Callin1}
\ba
  \lim_{y_r \to \infty} V(r) \eq \nonumber \\
  && \hspace{-18mm} -\frac{G m_1 m_2}{r} \left\{
    1 + \frac{8}{3\pi^2} \int_0^\infty \!\!
      \frac{e^{-mr}}{J_1^2(\frac{m}{\mu}) + Y_1^2(\frac{m}{\mu})} \frac{dm}{m}
  \right\} . \hspace{7mm}
\ea
This expression can be expanded to high order in powers of $\mu r$ applicable
to the limits $\mu r \gg 1$ and $\mu r \ll 1$~\cite{Callin1}.

From (\ref{eq:RS_pot_y}) we also have the potential for the more general case
when the two masses have different $y$-coordinates. For instance, one can
consider one mass on the visible brane at $y=0$ and the other mass on the
hidden brane at $y=y_r$. However, the interpretation of the resulting potential
is not obvious. This question has recently been addressed in~\cite{Arnowitt}.

\section{Summary}
\label{sec:summary}

In this paper we have derived the effective four-dimensional Lagrangian of
gravity in two different higher-dimensional theories, first using torus
compactification and then in the Randall-Sundrum braneworld model, and we have
used the result to calculate the Newtonian gravitational potential in the two
models. In both cases we have seen that a radion -- a massless scalar field --
contributes to the potential. Since the radion is massless, this affects the
long range gravitational force and therefore also the effective
four-dimensional Planck mass.

The presence of the radion is a result of the finite size of the fifth
dimension, and basically represents fluctuations in the circumference $L$ of
the torus and in the brane separation~$y_r$. In this paper we have simply
ignored the question of stabilizing the extra dimension, and the radion is
therefore massless. In a more realistic description, where the stabilization is
achieved by a dynamical mechanism (see e.g. \cite{GoldbergerWise}), the radion
usually becomes massive and will therefore not contribute to the long range
gravitational force.

In our derivation for the torus, we have found that a massless vector field
$A^\mu$ is part of the effective four-dimensional theory. This field does not
contribute to the gravitational interaction when the source is at rest in the
extra dimension. However, it still is a physical degree of freedom, and with a
moving source it can potentially contribute. In the RS model, on the other
hand, the additional orbifold symmetry effectively removes the vector field as
a physical degree of freedom. It is then a pure gauge mode that is removed by a
suitable choice of coordinates, which among other things requires that the
visible brane remains unbent in position $y=0$. The hidden brane is still
allowed to move, and the physical distance between the branes is precisely the
radion field $\phi(x)$ which cannot be gauged away.

\vspace{4mm}\textbf{Acknowledgement:} We want to thank David Langlois for
useful discussions about the role of the scalar field in these kinds of models.
This work has been supported by grant no. NFR 153577/432 and 159637/V30 from
the Research Council of Norway.

\appendix

\section{Metric perturbations in a conformally flat space}
\label{sec:conform_Lagr}

We consider an arbitrary conformally flat space with $D$ dimensions, where the
perturbed metric can be written
\be
  g_{\mu\nu}(x) =
    A^2(x) \left[ \addsmsp{ \eta_{\mu\nu} + h_{\mu\nu}(x) } \right] .
\ee
Here the warp factor $A(x)$ is allowed to depend on all coordinates. The
inverse metric is
\be
  g^{\mu\nu} = A^{-2} \left[ \addsmsp{
    \eta^{\mu\nu} - h^{\mu\nu} + h^{\mu\sigma} {h_\sigma}^\nu + \ldots
  } \right] .
\ee
The Lagrangian of the graviton field $h_{\mu\nu}(x)$ is found by expanding the
action
\be
  S = -\frac{1}{2} M^{D-2} \int d^D x \sqrt{g} R
\ee
to the second order in~$h_{\mu\nu}$. Of course, some other terms, like a
cosmological constant, must also be included in order to get a non-trivial warp
factor, but the curvature term is by far the hardest one to calculate. The
determinant of the metric is expanded as
\ba
  \sqrt{g} \eq A^D \sqrt{\det(\eta_{\mu\nu} + h_{\mu\nu})} \nonumber \\
  \eq A^D \cdot \exp \left[
    \frac{1}{2} \mathrm{Tr}
      \sum_{n=1}^\infty \frac{(-1)^{n+1}}{n} ({h^\mu}_\nu)^n
  \right] \nonumber \\
  \eq A^D \cdot \exp \left[
    \frac{1}{2} h - \frac{1}{4} h^{\alpha\beta} h_{\alpha\beta} + \ldots
  \right] \nonumber \\
  \eq A^D \left[
    1 + \frac{1}{2} h
    + \frac{1}{8} h^2 - \frac{1}{4} h^{\alpha\beta} h_{\alpha\beta}
    + \ldots
  \right] . \hspace{6mm}
  \label{eq:determinant}
\ea
The Christoffel symbols are, to the second order,
\ba
  {\Gamma^\mu}_{\alpha\beta} \eq
    \frac{A_\alpha}{A} \delta^\mu_\beta
    + \frac{A_\beta}{A} \delta^\mu_\alpha
    - \frac{A^\mu}{A} \eta_{\alpha\beta} \nonumber \\
  && \hspace{-10mm}
    + \, \frac{A_\nu}{A} \eta_{\alpha\beta} h^{\mu\nu}
    - \frac{A^\mu}{A} h_{\alpha\beta}
    + \frac{1}{2} \left(
      {h^\mu}_{\alpha,\beta}
      + {h^\mu}_{\beta,\alpha}
      - {h_{\alpha\beta}}^{,\mu}
    \right) \nonumber \\
  && \hspace{-10mm}
    + \, \frac{A_\nu}{A} h^{\mu\nu} h_{\alpha\beta}
    - \frac{A_\nu}{A} \eta_{\alpha\beta} h^{\mu\sigma} {h_\sigma}^\nu
    \nonumber \\
  && \hspace{-10mm}
    - \, \frac{1}{2} h^{\mu\nu} \left(
      h_{\nu\alpha,\beta} + h_{\nu\beta,\alpha} - h_{\alpha\beta,\nu}
    \right) ,
\ea
where $A_\mu \equiv \partial_\mu A$. The Ricci tensor is then found to be
\begin{widetext} \vspace{-5mm}
\ba
  R_{\mu\nu} \eq
    - \, (D-2) \frac{A_{\mu\nu}}{A}
    - \frac{A^\alpha_{\;\alpha}}{A}\eta_{\mu\nu}
    + (2D-4) \frac{A_\mu A_\nu}{A^2}
    - (D-3) \frac{A^\alpha A_\alpha}{A^2} \eta_{\mu\nu} \nonumber \\
  \eqnl + \, \frac{A_{\sigma\alpha}}{A} \eta_{\mu\nu} h^{\sigma\alpha}
    - \frac{A^\alpha_{\;\alpha}}{A} h_{\mu\nu}
    + (D-3) \frac{A_\sigma A_\alpha}{A^2} \eta_{\mu\nu} h^{\sigma\alpha}
    - (D-3) \frac{A^\alpha A_\alpha}{A^2} h_{\mu\nu} \nonumber \\
  \eqnl + \, (D-2) \frac{A^\sigma}{2A} \left(
      h_{\sigma\mu,\nu} + h_{\sigma\nu,\mu} - h_{\mu\nu,\sigma}
    \right) + \frac{1}{2} \left(
      {h^\alpha}_{\mu,\nu\alpha} + {h^\alpha}_{\nu,\mu\alpha}
      - h_{,\mu\nu} - \Box h_{\mu\nu}
    \right) + \frac{A^\sigma}{2A} \eta_{\mu\nu} \left(
      2 {h^\alpha}_{\sigma,\alpha} - h_{,\sigma}
    \right) \nonumber \\
  \eqnl + \, \frac{A_{\sigma\alpha}}{A} h^{\sigma\alpha} h_{\mu\nu}
    - \frac{A_{\sigma\alpha}}{A} \eta_{\mu\nu} h^{\alpha\rho} {h_\rho}^\sigma
    + (D-3) \frac{A_\sigma A_\alpha}{A^2} h^{\sigma\alpha} h_{\mu\nu}
    - (D-3) \frac{A_\sigma A_\alpha}{A^2}
         \eta_{\mu\nu} h^{\alpha\rho} {h_\rho}^\sigma \nonumber \\
  \eqnl + \, \frac{A^\rho}{A} \eta_{\mu\nu} h^{\alpha\sigma} \left(
      \frac{1}{2} h_{\alpha\sigma,\rho} - h_{\rho\alpha,\sigma}
    \right)
    + \frac{A_\sigma}{A} \eta_{\mu\nu} h^{\rho\sigma} \left(
      \frac{1}{2} h_{,\rho} - {h^\alpha}_{\rho,\alpha}
    \right)
    + \frac{A^\sigma}{A} h_{\mu\nu} \left(
      {h^\alpha}_{\sigma,\alpha} - \frac{1}{2} h_{,\sigma}
    \right) \nonumber \\
  \eqnl - \, \frac{D-2}{2} \frac{A_\sigma}{A} h^{\rho\sigma} \left(
      h_{\rho\mu,\nu} + h_{\rho\nu,\mu} - h_{\mu\nu,\rho}
    \right)
    + \frac{1}{4} \left(
      h_{\sigma\mu,\nu} + h_{\sigma\nu,\mu} - h_{\mu\nu,\sigma}
    \right) \left(
      h^{,\sigma} - 2 {h^{\alpha\sigma}}_{,\alpha}
    \right) \nonumber \\
  \eqnl + \, \frac{1}{4} h_{\sigma\alpha,\mu} {h^{\sigma\alpha}}_{,\nu}
    + \frac{1}{2} h_{\sigma\mu,\alpha} {h_\nu}^{\sigma,\alpha}
    - \frac{1}{2} h_{\sigma\mu,\alpha} {h_\nu}^{\alpha,\sigma}
    + \frac{1}{2} h^{\alpha\sigma} \left(
      h_{\mu\nu,\alpha\sigma} + h_{\alpha\sigma,\mu\nu}
      - h_{\alpha\mu,\nu\sigma} - h_{\alpha\nu,\mu\sigma}
    \right) ,
\ea
and the curvature scalar is therefore
\ba
  A^2 R \eq
    - \, 2(D-1) \frac{A^\alpha_{\;\alpha}}{A}
    - (D-1)(D-4) \frac{A^\alpha A_\alpha}{A^2}
    + 2(D-1) \frac{A_{\alpha\beta}}{A} h^{\alpha\beta} \nonumber \\
  \eqnl + \, (D-1)(D-4) \frac{A_\alpha A_\beta}{A^2} h^{\alpha\beta}
    + 2(D-1) \frac{A_\sigma}{A} {h^{\sigma\alpha}}_{,\alpha}
    - (D-1) \frac{A^\sigma}{A} h_{,\sigma}
    + {h^{\alpha\beta}}_{,\alpha\beta} - \Box h \nonumber \\
  \eqnl - \, 2(D-1) \frac{A_{\alpha\beta}}{A} h^{\alpha\sigma} {h_\sigma}^\beta
    - (D-1)(D-4) \frac{A_\alpha A_\beta}{A^2} h^{\alpha\sigma} {h_\sigma}^\beta
    + (D-1)\frac{A^\sigma}{A} h^{\alpha\beta} h_{\alpha\beta,\sigma}
    \label{eq:R_conform} \\
  \eqnl - \, 2(D-1) \frac{A^\sigma}{A} h^{\alpha\beta} h_{\sigma\alpha,\beta}
    - 2(D-1) \frac{A_\sigma}{A} h^{\nu\sigma} {h^\alpha}_{\nu,\alpha}
    + (D-1) \frac{A_\sigma}{A} h^{\nu\sigma} h_{,\nu} \nonumber \\
  \eqnl + \, \frac{3}{4} h_{\alpha\beta,\nu} h^{\alpha\beta,\nu}
    - \frac{1}{2} h_{\alpha\beta,\nu} h^{\nu\beta,\alpha}
    + {h^{\alpha\beta}}_{,\beta} h_{,\alpha}
    - {h^{\alpha\nu}}_{,\alpha} {h^\beta}_{\nu,\beta}
    - \frac{1}{4} h_{,\alpha} h^{,\alpha}
    + h^{\alpha\beta} h_{,\alpha\beta}
    + h^{\alpha\beta} \Box h_{\alpha\beta}
    - 2 {h_\alpha}^\beta {h^{\alpha\nu}}_{,\nu\beta} \, . \nonumber
\ea
Combining (\ref{eq:determinant}) and (\ref{eq:R_conform}) we get the rather
complicated expression
\ba
  A^{2-D} \sqrt{g} R \eq
    - \, 2(D-1) \frac{A^\alpha_{\;\alpha}}{A}
    - (D-1)(D-4) \frac{A^\alpha A_\alpha}{A^2}
    + 2(D-1) \frac{A_{\alpha\beta}}{A} h^{\alpha\beta}
    + (D-1)(D-4) \frac{A_\alpha A_\beta}{A^2} h^{\alpha\beta} \nonumber \\
  \eqnl - \, (D-1) \frac{A^\alpha_{\;\alpha}}{A} h
    + 2(D-1) \frac{A_\sigma}{A} {h^{\sigma\alpha}}_{,\alpha}
    - (D-1) \frac{A^\sigma}{A} h_{,\sigma}
    - \frac{1}{2}(D-1)(D-4) \frac{A^\alpha A_\alpha}{A^2} h
    + {h^{\alpha\beta}}_{,\alpha\beta} - \Box h \nonumber \\
  \eqnl - \, 2(D-1) \frac{A_{\alpha\beta}}{A} h^{\alpha\sigma} {h_\sigma}^\beta
    - (D-1)(D-4) \frac{A_\alpha A_\beta}{A^2} h^{\alpha\sigma} {h_\sigma}^\beta
    + (D-1) \frac{A^\sigma}{A} h^{\alpha\beta} h_{\alpha\beta,\sigma}
    \nonumber \\
  \eqnl - \, 2(D-1) \frac{A^\sigma}{A} h^{\alpha\beta} h_{\sigma\alpha,\beta}
    - 2(D-1) \frac{A_\sigma}{A} h^{\nu\sigma} {h^\alpha}_{\nu,\alpha}
    + (D-1) \frac{A_\sigma}{A} h^{\nu\sigma} h_{,\nu}
    + (D-1) \frac{A_{\alpha\beta}}{A} h^{\alpha\beta} h \nonumber \\
  \eqnl + \, \frac{1}{2}(D-1)(D-4) \frac{A_\alpha A_\beta}{A^2} h^{\alpha\beta} h
    + (D-1) \frac{A_\sigma}{A} {h^{\sigma\alpha}}_{,\alpha} h
    - \frac{D-1}{2} \frac{A^\sigma}{A} h_{,\sigma} h
    - \frac{D-1}{4} \frac{A^\alpha_{\;\alpha}}{A} h^2 \nonumber \\
  \eqnl - \, \frac{1}{8}(D-1)(D-4) \frac{A^\alpha A_\alpha}{A^2} h^2
    + \frac{D-1}{2} \frac{A^\alpha_{\;\alpha}}{A} h^{\mu\nu} h_{\mu\nu}
    + \frac{1}{4}(D-1)(D-4) \frac{A^\alpha A_\alpha}{A^2} h^{\mu\nu} h_{\mu\nu}
    \nonumber \\
  \eqnl + \, \frac{3}{4} h_{\alpha\beta,\nu} h^{\alpha\beta,\nu}
    - \frac{1}{2} h_{\alpha\beta,\nu} h^{\nu\beta,\alpha}
    + {h^{\alpha\beta}}_{,\beta} h_{,\alpha}
    - {h^{\alpha\nu}}_{,\alpha} {h^\beta}_{\nu,\beta}
    - \frac{1}{4} h_{,\alpha} h^{,\alpha} \nonumber \\
  \eqnl + \, h^{\alpha\beta} h_{,\alpha\beta}
    + h^{\alpha\beta} \Box h_{\alpha\beta}
    - 2 {h_\alpha}^\beta {h^{\alpha\nu}}_{,\nu\beta}
    + \frac{1}{2} {h^{\alpha\beta}}_{,\alpha\beta} h
    - \frac{1}{2} h \Box h \; .
\ea
When inserted into the integral of the action, the zeroth order and first order
parts of this expression should always vanish when the warp factor $A(x)$ is a
solution to the background field equations. The second order part can be
simplified using partial integrations, with the result
\ba
  S \eq -\frac{1}{2} M^{D-2} \int d^D x \left\{
    A^{D-4} \left[
      (D-1)(D-2) A_\alpha A_\beta h^{\alpha\sigma} {h_\sigma}^\beta
      - \frac{1}{4}(D-1)(D-2) A_\sigma A^\sigma h^{\alpha\beta} h_{\alpha\beta}
    \right.
  \right. \nonumber \\
  && \hspace{39mm} \left.
    \left.
      - \, \frac{1}{2}(D-1)(D-2) A_\alpha A_\beta h^{\alpha\beta} h
      + \frac{1}{8}(D-1)(D-2) A_\alpha A^\alpha h^2
    \right]
  \right. \nonumber \\
  && \hspace{27.9mm} \left.
    + \, (D-2) A^{D-3} \left[
      \frac{1}{2} A^\alpha h_{\alpha\beta,\nu} h^{\nu\beta}
      - A_\alpha h^{\alpha\beta} h_{,\beta}
      - A^\nu h^{\alpha\beta} h_{\alpha\beta,\nu}
    \right.
  \right. \nonumber \\
  && \hspace{54.2mm} \left.
    \left.
      + \, \frac{3}{2} A_\beta {h_\alpha}^\beta {h^{\alpha\nu}}_{,\nu}
      - \frac{1}{2} A_\beta {h^{\alpha\beta}}_{,\alpha} h
      + \frac{1}{2} A^\alpha h h_{,\alpha}
    \right]
  \right. \nonumber \\
  && \hspace{27.9mm} \left.
    + \, A^{D-2} \left[
      - \, \frac{1}{4} h_{\alpha\beta,\nu} h^{\alpha\beta,\nu}
      - \frac{1}{2} {h^{\alpha\beta}}_{,\beta} h_{,\alpha}
      + \frac{1}{2} {h^{\alpha\nu}}_{,\alpha} {h^\beta}_{\nu,\beta}
      + \frac{1}{4} h_{,\alpha} h^{,\alpha}
    \right]
  \right\} .
  \label{eq:action_general}
\ea
This result can then be applied to e.g. the RS model, provided we also include
the cosmological constant and brane tension. With $A(x) = 1$ we see that
(\ref{eq:action_general}) is trivially reduced to the flat space
result~(\ref{eq:graviton_Lagr}). \vspace{5mm}
\end{widetext}


\begin{thebibliography}{10}
\bibitem{footnote_potential}   
In general, we define the gravitational constant $G_D$ in $D$ spacetime
dimensions so that the gravitational force is given by $F \equiv G_D m_1 m_2 /
r^{D-2} = dV/dr$, meaning that $V(r) = -G_D m_1 m_2 / \left[ (D-3) \, r^{D-3}
\right]$.
\end{thebibliography}
\end{document}